\documentclass[aps,prl,reprint,superscriptaddress,nobalancelastpage]{revtex4-2}
\usepackage{graphicx,amsmath,amssymb,amsthm,xcolor}
\usepackage{hyperref}
\usepackage{bm}
\DeclareMathAlphabet{\mathcal}{OMS}{cmsy}{m}{n}
\definecolor{linkcolor}{HTML}{0000ff}
\hypersetup{colorlinks,linkcolor={linkcolor},citecolor={linkcolor},urlcolor={linkcolor}}

\begin{document}

\title{A Constrained Mechanical Metamaterial Towards Wave Polarization and Steering Control}

\author{Shiheng Zhao}
\altaffiliation{Current address: Max Planck Institute for the Physics of Complex Systems, N\"othnitzer Stra\ss e 38, 01187 Dresden, Germany}
\affiliation{College of Science, China Agricultural University, Beijing 100083, China}
\author{Zhan Tian}
\affiliation{College of Science, China Agricultural University, Beijing 100083, China}
\author{Jiaji Chen}
\affiliation{Department of Mechanical and Aerospace Engineering, University of Missouri, Columbia, Missouri 65211, USA}
\author{Heng Jiang}
\affiliation{Key Laboratory of Microgravity, Institue of Mechanics, Chinese Academy of Sciences, Beijing 100190, China}
\author{Zheng Chang}
\altaffiliation{Z. Chang: changzh@cau.edu.cn}
\affiliation{College of Science, China Agricultural University, Beijing 100083, China}
\author{Guoliang Huang}
\altaffiliation{G. Huang: guohuang@pku.edu.cn;huangg@missouri.edu.}
\affiliation{Department of Mechanical and Aerospace Engineering, University of Missouri, Columbia, Missouri 65211, USA}
\affiliation{Department of Mechanics and Engineering Science, College of Engineering, Peking University, Beijing 100871, PR China}

\date{\today}

\begin{abstract}
Precise control of the polarization and propagation direction of elastic waves is a fundamental challenge in elastodynamics. Achieving efficient mode conversion along arbitrary paths with conventional techniques has proven difficult. In this letter, we propose an innovative harmonimode mechanical metamaterial by integrating classical lattice architecture with a constrained mechanism. The constrained discrete mass-spring model is formulated and homogenized to reveal the unique harmonimode behavior, which supports single-mode polarized propagation and perfect impedance matching with the reference medium. Leveraging multi-scale simulations and the discrete transformation method, the metamaterial is designed to exhibit degenerated wave polarization and broadband mode conversion along various paths by simply adjusting constraint orientations. Finally, hinge joints are proposed for the physical realization of the metamaterial with sub-wavelength microstructures. Numerical simulations confirm its exceptional wave control performance over a broad frequency range. This work presents a comprehensive framework for designing harmonimode metamaterials capable of arbitrary polarization control.

\end{abstract}

\maketitle

Manipulating the propagation of elastic waves in solid materials and structures is crucial for a wide range of applications, including medical imaging, structural health monitoring, signal processing, vibration suppression, and energy harvesting. Unlike electromagnetic and acoustic counterparts, elastic waves possess rich polarization characteristics, enabling advanced functionalities such as mode splitting and mode conversion~\cite{Auld1990}. Mechanical metamaterials~\cite{Craster2023}, as artificial structures with novel properties, offer significant potential in controlling elastic waves, allowing for precise steering~\cite{stenger2012experiments,Li2021MSSP} and polarization~\cite{liu2019designing}, both passively and actively. 

Wave steering is typically realized by designing the material behaviors through transformed elasticity~\cite{milton2006cloaking,Hu2011PRB,norris2011elastic,chang2011APL} such as cloaking~\cite{Farhat2009PRL,Li2021MSSP,Li2020IJSS}, while polarization control mainly relies on structural symmetry breaking or material anisotropy~\cite{chang2015APL,kweun2017transmodal,frenzel2019ultrasound,chen2020activeNC,cao2021JMPS,Dong2022PRApp,Wu2022Ijss}. Nevertheless, designing metamaterials for both wave steering and wave mode polarization remains elusive. Current research provides little insight into the underlying microstructure of these transformed materials, making their design a major challenge in the field.

Recently, discrete transformation elasticity (DTE)~\cite{chen2021discrete,zhao2023microstructure} has been developed as a promising solution for realize microstructures for arbitary wave path manipulation. A key advantage of DTE is in its ability to modulate mass and spring components independently, providing design flexibility and paving the way for advanced functionalities. For example, torsion springs on mass blocks can achieve the desired polar elastic tensor for cloaking while maintaining the structure's effective mass density ~\cite{nassar2018degenerate}.
This modulation capacity suggests that constraints could introduce novel functionalities, such as polarization control, without compromising DTE's path manipulation capabilities. However, the understanding of how these microscopic constraints influence macroscopic behavior remains limited. Many key questions need to be addressed: How can we accurately characterize the macroscopic behavior of constrained metamaterials? What unique features arise from varying constraints? How do these features translate into practical applications? 
In this letter, we propose an innovative mechanical metamaterial by harnessing classical lattice architecture with a constrained mechanism. Through theoretical analysis and multi-scale numerical simulations, we elucidate the elastodynamics of this material and its innovative wave functionalities for simultaneous wave polarization and steering. Finally, we propose a straightforward and practical approach for microstructure realization and demonstrate its excellent wave control performance.

\begin{figure*}[ht!]
    \centering
    \includegraphics[width = 0.8\linewidth]{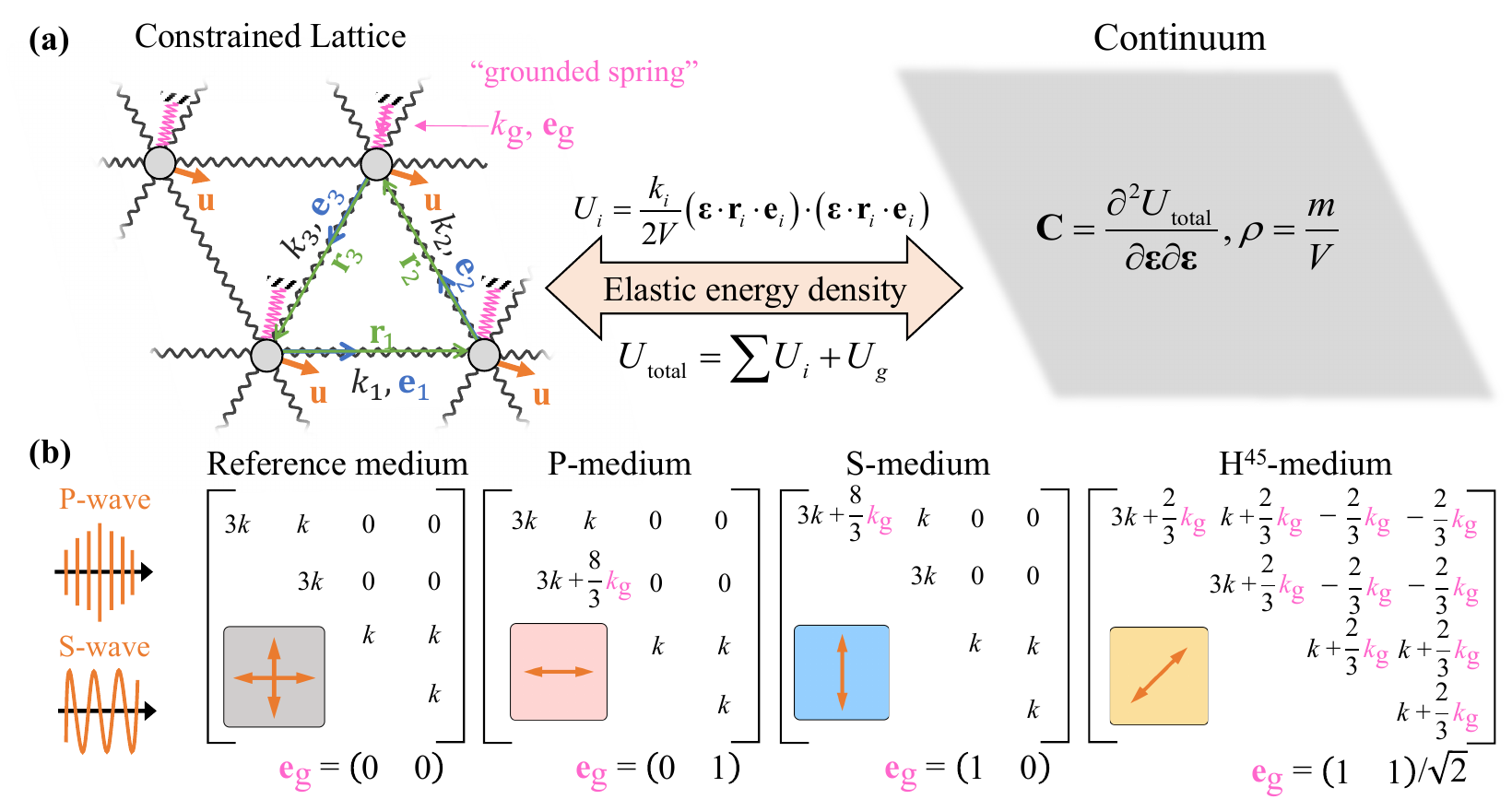}
    \caption{Theoretical Model and Elastic Tensors of Constrained Metamaterials. (a) A mass-spring lattice with a constrained design, where masses linked by springs are anchored by much stiffer grounded springs. This configuration, which limits certain movements, underpins the homogenized elastodynamics represented by an effective elastic tensor and mass density. (b) Normalized effective elastic tensors for different grounded spring orientations, categorizing the media into Reference, P-, S-, and H$^{45}$-media for a horizontal wave vector, based on their wave polarization capabilities, with each type marked by distinct icons for easy identification.}
    \label{fig1}
\end{figure*}

Consider the triangular lattice sketched in Fig.~\ref{fig1}(a) with uniform springs ($k_i=k$, where $i=1,2,3$). The spring orientations ($\mathbf{e}_i$) are determined by the lattice vectors ($\mathbf{r}_i$), such that $\mathbf{e}_i = \mathbf{r}_i/|\mathbf{r}_i|$ . This lattice serves as an isotropic reference for the constrained design~\cite{ashby1997cellular}. Each unit cell hosts a rigid mass constrained by an in-plane ground spring, which is significantly stiffer than the inter-mass springs ($k_g \gg k $). The ground springs restrict both rotational and one translational degree of freedom, allowing movement only in the direction perpendicular to their orientation. In general, such a medium can be classified as a reduced case of a polar lattice when fully constrained~\cite{nassar2018degenerate}. 
The homogenized elastodynamic behavior of the constrained metamaterial is characterized by its effective fourth-order elastic tensor $\mathbf{C}$:
\begin{equation}\label{eq1}
\mathbf{C} = \frac{\partial^2 U_{\text{total}}}{\partial \bm{\varepsilon} \partial \bm{\varepsilon}},
\end{equation}
where $U_{\text{total}} = \sum{U_j}$, with $j = 1, 2, 3,$ and $g$, represents the total elastic energy of the system, and $\bm{\varepsilon} = \nabla \mathbf{u} = \nabla (u_x, u_y)$ is the displacement gradient. The elastic energy $U_j$ for each spring type is uniformly calculated using the expression $U_j=k_j (\bm{\varepsilon} \cdot \mathbf{r}_j \cdot \mathbf{e}_j)^2/2V$. The isotropic effective mass density of the constrained metamaterial is simply given by $\rho = m / V$, where $m$ is the mass of the unit cell and $V=(\sqrt{3} |\mathbf{r}_i|^2)/2$ denotes the area of the lattice unit cell.
   
Fig.~\ref{fig1}(b) illustrates the normalized effective elastic tensors for various orientations of ground springs, scaled by $\sqrt{3}/4$. These tensors exhibit major symmetry, so only the upper triangular elements are shown. It is also noteworthy that minor symmetry is preserved, assuming the size and rotation of the masses are disregarded. The stiffness tensor of the reference medium, without ground springs ($\mathbf{e}_g=(0,0)$), exhibits isotropic properties with Lamé constants $\lambda=\mu=(\sqrt{3}k)/4$. Introducing ground springs with varying orientations modifies the effective elastic tensor, thereby altering the medium's wave propagation characteristics. For example, when the ground springs are oriented vertically ($\mathbf{e}_g=(0,1)$), the stiffness $k_g$ is associated with the $C_{2222}$ component. This indicates that for a horizontal wave vector, the medium permits the propagation of longitudinal (P-mode) waves, similar to the reference medium, while completely blocking their transverse (S-mode)  counterpart. We define this medium as a ``\textit{\textbf{harmonimode}}" medium because \textit{it perfectly matches the impedance of the reference medium (\textbf{harmonic}) and supports only a single propagation mode while suppressing all others (\textbf{omni-mode)}}. This unique characteristic is crucial for achieving precise mode separation and efficient elastic wave filtering. A medium that supports only P-mode is called a P-continuum or P-lattice. Conversely, a horizontal orientation of the ground spring ($\mathbf{e}_g=(1,0)$) results in an S-medium. The final panel of Fig.~\ref{fig1}(b) shows the effective elastic tensor with the ground springs oriented at 45° ($\mathbf{e}_g=(\sqrt{2}/2,\sqrt{2}/2)$). This medium permits 45°-polarized waves but does not exhibit harmonimode behavior for horizontal waves, classifying it as a Hybrid-45 (H$^{45}$) medium. 

\begin{figure}[t]
    \centering
    \includegraphics[width=\linewidth]{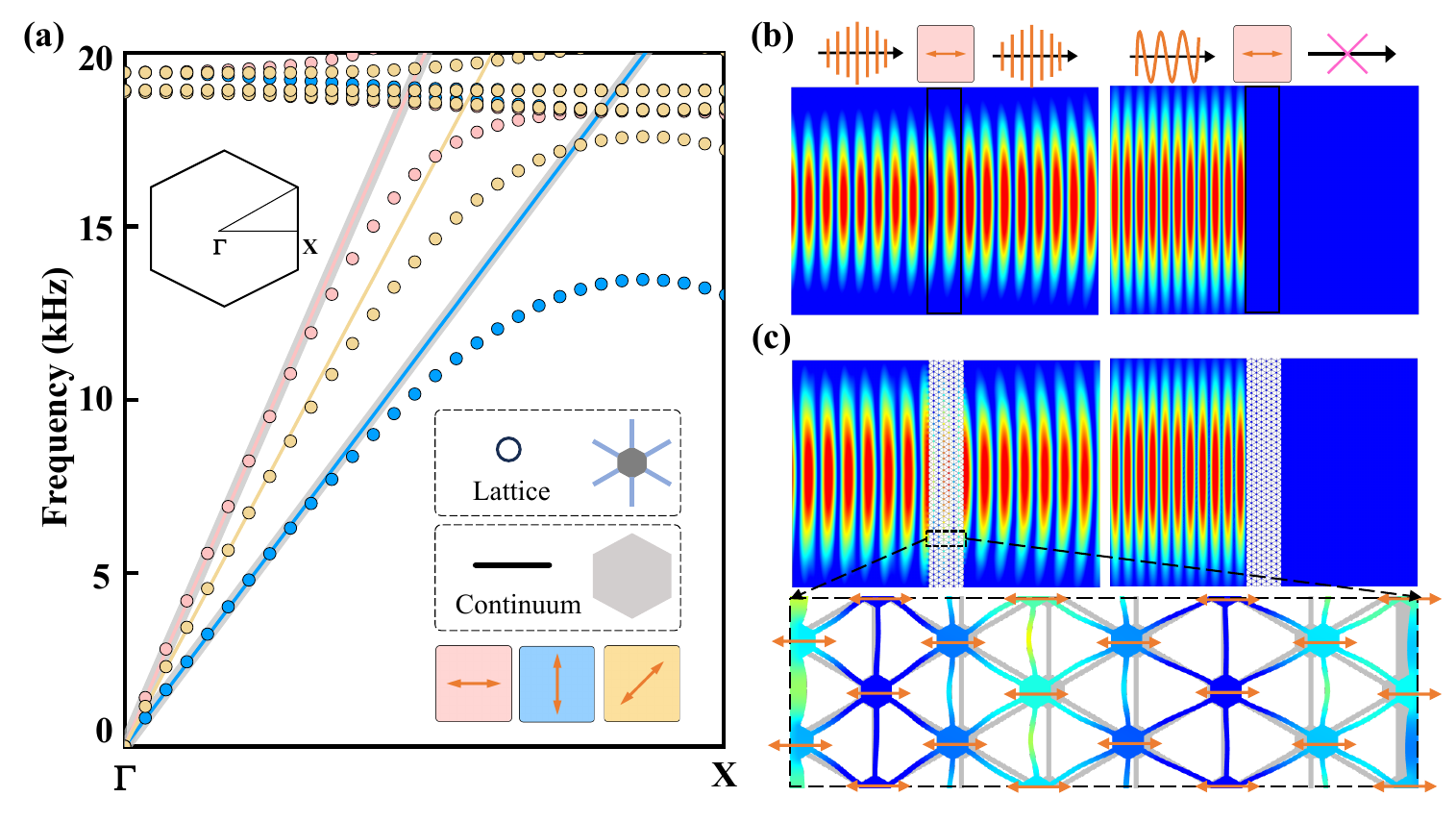}
    \caption{Elastodynamics of Homogeneous Constrained Media. (a) Dispersion relations in the $\mathrm{\Gamma}-X$ direction of the first irreducible Brillouin zone, contrasting the continuum (solid lines) and lattice structures (scatter plots), with each media type color-coded. (b) and (c) show wave field simulations confirming the harmonimode behavior of the P-media, as evidenced by the total displacement field for specific P- and S-wave frequencies.}
    \label{fig2}
\end{figure}

To validate the wave polarization control capabilities of the constrained medium, 2D numerical simulations were performed on both the continuum medium with effective stiffness and the microstructured lattices. Polyoxymethylene (POM,~\cite{ashby2007teaching}), a common polymer, was chosen as the reference continuous medium, with a Young's modulus of $E_c =1.4$ GPa, Poisson's ratio $\nu_c = 0.33$, and mass density $\rho_c = 1350$ kg/m$^3$. This medium was discretized into a lattice with stiffness $k = 1.21$ GPa. The stiffness of the constraint $k_g$ was set to $10^8$ times $k$, providing an extremely firm, though not perfectly rigid, constraint. To address practical challenges, the microstructure shown in the top panel of the inset in Fig.~\ref{fig2}(a) was realized, with each unit cell consisting of six slender rods and a dodecagonal mass block (Supplemental Material, Sec. I \footnote{See Supplemental Material at [url to be inserted], for (i) the material and structure of the 2D constrained lattice, (ii) slowness curves of constrained media, (iii) reflection and refraction at the interface, (iv) geometric insensitivity in harmonimode propagation, (v) minimal structure for mode conversion, (vi) non-reciprocity and asymmetric propagation, (vii) kinetic energy equivalence in 3D structure design, and (viii) supplementary information on 3D simulation analysis.}).

Fig.~\ref{fig2}(a) shows the dispersion relations obtained using the solid-mechanics module of COMSOL Multiphysics, comparing the continuum medium with the corresponding microstructured lattices to assess their elastodynamic equivalence and determine the operational frequency band. In this figure, solid lines represent the dispersion properties of the continuum, while scatter plots depict those of the microstructured lattices. The reference continuum exhibits conventional non-dispersive propagation in both P- and S-modes. In contrast, the P- and S-continua exhibit distinctive harmonimode behavior, each supporting a single branch that precisely aligns with the corresponding mode of the reference medium. As expected, the dispersion curve of the H$^{45}$-medium falls between those of the P- and S-media. The constrained lattices replicate the harmonimode behavior observed in the continuum at low frequencies but deviate as frequency increases due to the long-wave approximation. Within the 0–15 kHz and 0–10 kHz ranges, respectively, the P- and S-lattices align closely with their corresponding continua. 

To further validate harmonimode behavior, we conducted frequency-domain wave propagation simulations using a reference medium sandwiched with the constrained medium, both in continuum and discretized lattice forms. Fig.~\ref{fig2}(b) and (c) show Gaussian wave beams of 10 kHz P-wave and 8 kHz S-wave propagating horizontally through a reference continuum embedded with P-continuum and a P-lattice of $7\times40$ unit cells, respectively. The results confirm the efficiency of mode filtering: both the P-continuum and P-lattice allow unimpeded transmission of P-waves while effectively blocking S-waves. A close-up of the P-lattice in the bottom panel of Fig.~\ref{fig2}(c) further supports this selective propagation, showing all mass blocks moving strictly in alignment with the horizontal direction of the P-wave. To fully understand the elastic wave behavior in the constrained media, we computed their slowness curves and analyzed refraction and reflection at the interfaces with the reference media (Supplemental Material, Sec. II \& III ~\cite{Note1}).  Notably, the harmonimode behavior is preserved across any geometric interface between the constrained and reference media, enabling a wide range of applications (Supplemental Material, Sec. IV ~\cite{Note1}).

\begin{figure}[t]
    \centering
    \includegraphics[width=\linewidth]{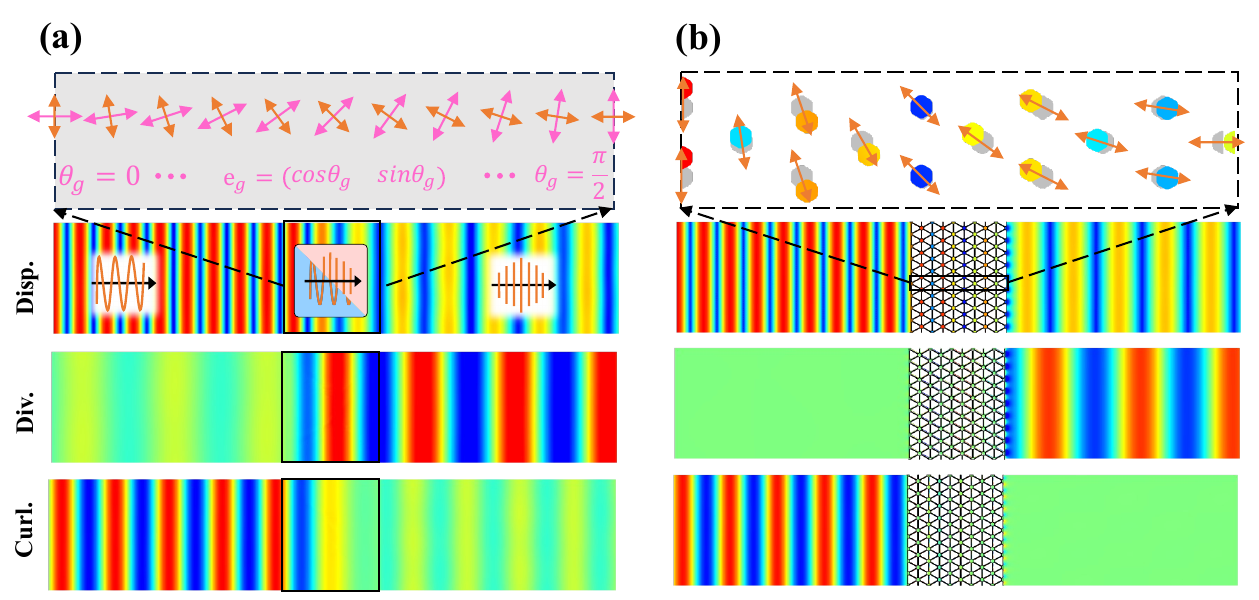}
    \caption{Mode Conversion in Gradient Constrained Media. (a) Demonstrates S-to-P wave conversion in a continuum medium, with constraint orientations transitioning linearly from 0 to $\pi/2$ across a $9 \times 9$ unit cell domain. The total displacement, divergence, and curl fields validate the conversion. (b) Depicts mode conversion in the lattice medium, evidenced by the displacement modes of mass blocks, confirming the effective transformation of an S-wave into a P-wave.}
    \label{fig3}
\end{figure}

To achieve wave mode conversion, a 1D constrained medium with gradient orientation is applied: in the continuum, it spans 9 discrete unit cells in length, and in the discrete lattice, it forms a 9-unit cell grid, as shown in Fig.~\ref{fig3} (a) and (b). In the study, we apply Bloch-Floquet conditions on the upper and lower boundaries to simulate an infinite vertical extent. The constrain variation follows a linear gradient along the horizontal axis, shifting from 0 to $\pi/2$. This gradient transforms the medium from an S-medium on the left to a P-medium on the right, ensuring each end is ``harmonimode" to the reference medium. As shown in Fig.~\ref{fig3}, the incident S-mode wave is seamlessly converted into a P-mode wave, which is observable in the total displacement field and even more pronounced in the divergence and curl fields, providing nuanced insights into the wave conversion process. In Fig.\ref{fig3}(a), minor reflections and transmissions are observed in the divergence and curl fields, likely due to the use of approximate rigid constraints in the continuous model. In contrast, Fig.\ref{fig3}(b) shows cleaner reflection and transmission fields, as the rigid constraints are directly applied to the mass blocks in the lattice model. An in-depth observation of how the lattice’s mass blocks respond to wave incidence is provided, with grey indicating reference positions and colored representations showing mass motion. Notably, this design allows efficient mode conversion with just a two unit-cell-thickness constrained lattice (Supplemental Material, Sec. V ~\cite{Note1}). In addition, the designed medium is inherently non-reciprocal. For instance, an S-wave traveling from right to left is entirely blocked on the right side of the medium, and vice versa. (Supplemental Material, Sec. VI ~\cite{Note1}). 

\begin{figure}[t]
    \centering
    \includegraphics[width=0.9\linewidth]{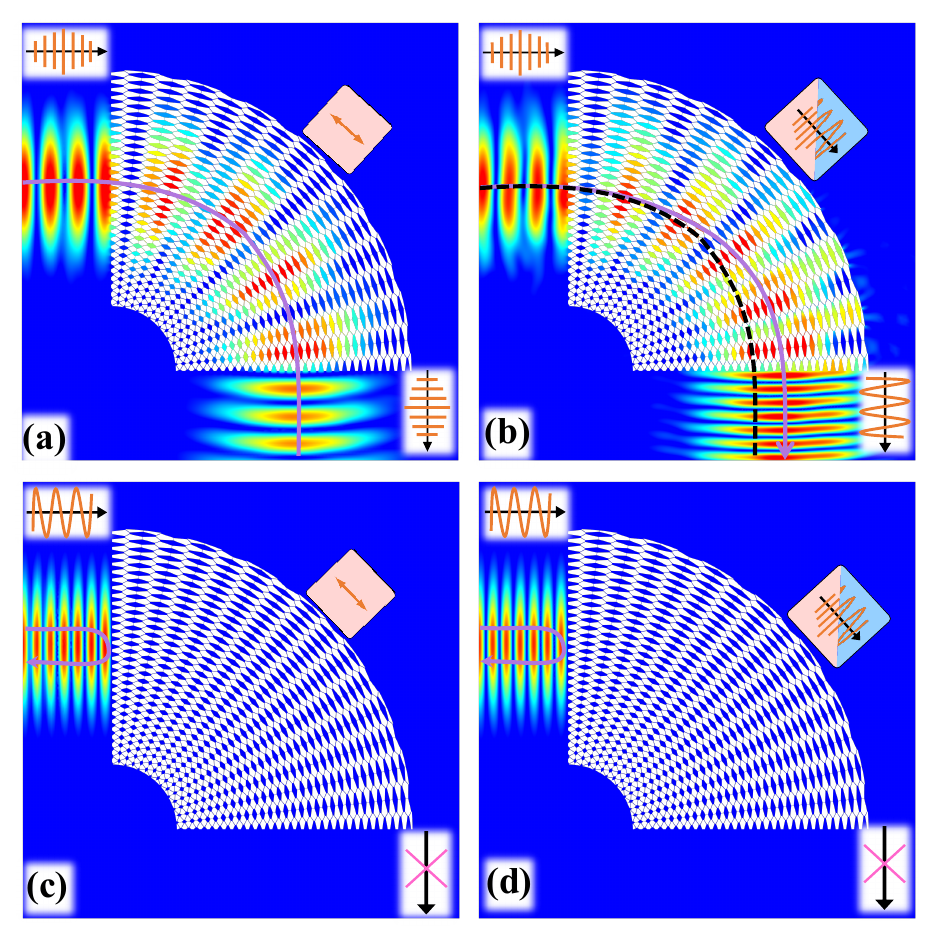}
    \caption{Functional Bending Waveguides in Constrained Media. (a) A P-wave navigating through a harmonimode waveguide, with auxiliary curves (purple) tracing its trajectory. (b) Mode conversion of a P-wave, highlighted by the deviation from the waveguide centerline (black). (c) and (d) depict analogous scenarios for an S-wave in harmonimode and mode conversion waveguides.}
    \label{fig4}
\end{figure}

Building on recent advancements, we propose using the constrained medium with DTE to achieve, for the first time, simultaneous control of both wave steering and mode conversion.
Specifically, two distinct types of bending waveguides for perfect mode filtering and conversion are designed and numerically tested, as illustrated in Fig.~\ref{fig4}. In the simulation, microstructures of the transformation medium are obtained using DTE ~\cite{zhao2023microstructure} on polar-lattice waveguides. We then precisely constrain each mass block to transform the medium into either a harmonimode or a mode conversion medium. For the harmonimode medium, constraints are arranged parallel or perpendicular to the path to maintain a consistent wave flow. Conversely, in the mode conversion medium, the constrain orientations undergo a gradual shift from parallel (or perpendicular) at the entry to perpendicular (or parallel) at the exit, enabling a progressively wave transformation process. Fig.\ref{fig4}(a) shows a P-wave beam smoothly navigating a harmonimode bending waveguide, achieving a precise 90° turn. Conversely, Fig.\ref{fig4}(b) demonstrates the successful conversion of a P-wave into an S-wave, as evidenced by the wavelength change in the outgoing wave. Gradient variation in constraints also leads to asymmetric wave propagation (Supplemental Material, Sec. VI ~\cite{Note1}), observable as the beam deviates from the waveguide's centerline. Unlike media based on global transformation theory, the constrained media support only a single mode of wave propagation, such as P-mode or S-mode. As shown in Figs.~\ref{fig4}(c) and (d), the current waveguides are single P-mode media, effectively blocking S-mode wave propagation. 

\begin{figure}[t]
    \centering
    \includegraphics[width=\linewidth]{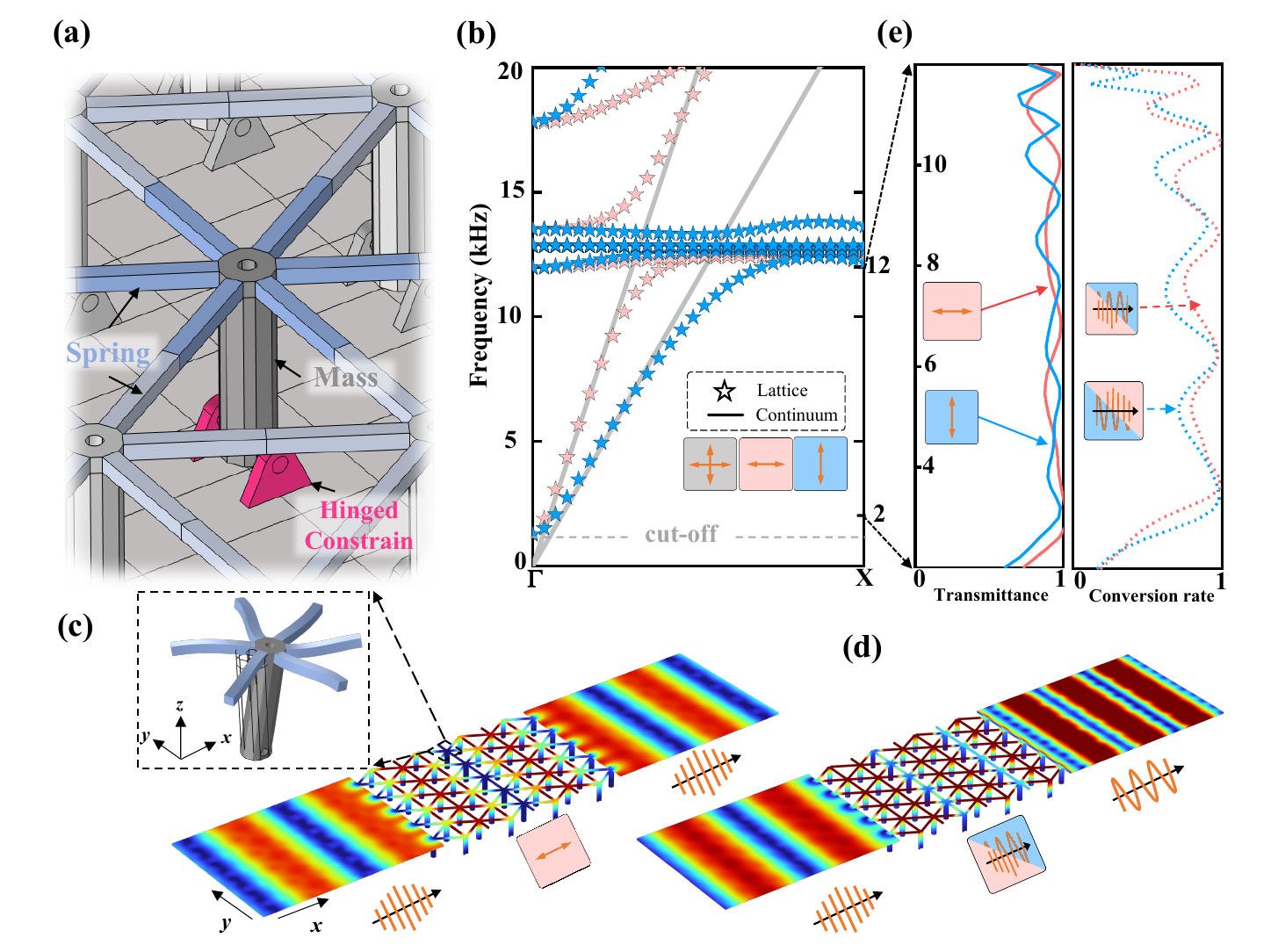}
    \caption{Structural Implementation of Constrained Media. (a) A unit cell design featuring a hexagonal prism with hinge constraints, representing a constrained mass block. (b) The dispersion relation, showing harmonimode behavior between 3-10 kHz, with resonances near 12.5 kHz and a cutoff around 1 kHz. (c) and (d) Wavefield simulations for P-medium and P-to-S medium, respectively, demonstrating harmonimode transmission and mode conversion at 10 kHz. (e) The transmission spectrum, highlighting the robust performance of the constrained structures across a broad frequency range of 2-10 kHz.}
    \label{fig5}
\end{figure}

Achieving effective constrained elastic behavior is crucial for the practical implementation of the proposed medium. In this study, we introduce a novel design method for mechanical constraints and evaluate these structures numerically for potential wave control applications. Fig.~\ref{fig5}(a) depicts the unique constraint mechanism, consisting of a hexagonal prism hinged and grounded at one end, with six rods connected at the opposite end, following the previous 2D lattice configuration (Supplemental Material, Sec. VII ~\cite{Note1}). A key aspect of the design is integrating constraints with mass-block movements, based on the principles of kinetic energy equivalence. The flexibility of this design allows for adjusting constraint directions, making it adaptable to specific functional requirements.

To validate our design, the dispersion relations for the 3D constrained structure are calculated for both the P-lattice and S-lattice, as shown in Fig.~\ref{fig5}(b), and compared with those from the effective continuum model. In the simulation, hinge joints are used to accurately model smooth hinge constraints within the structure (Supplemental Material, Sec. VIII ~\cite{Note1}). As shown in the figure, despite resonances near 13 kHz and a grounding-induced cutoff around 1 kHz, our design successfully achieves harmonimode equivalence with the reference medium across a broad frequency range of 3-10 kHz for both the P-lattice and S-lattice. 

Full-scale wavefield simulations were further conducted to examine the performance of the constrained structures, focusing on the P-lattice (Fig.\ref{fig5}(c)) and the P-to-S lattice with gradient orientations (Fig.\ref{fig5}(d)). To optimize computational efficiency, we used eight unit cells along the $x$-direction and six along the $y$-direction. For a P-wave propagating at 9 kHz along the $x$-axis, the designed lattices successfully demonstrated harmonimode transmission in the P-medium, and effective mode conversion in the P-to-S medium. These results vividly showcase the practical effectiveness of our structural design. Additionally, the comprehensive wave transmission performance of the constrained structures was numerically evaluated across a wide frequency range, from 2 to 10 kHz, as depicted in Fig.~\ref{fig5}(e). In this analysis, we define transmittance as the ratio of the output wave energy to the input ($T_{tr} = W_{\text{out}} / W_{\text{in}})$, where $W_{\text{in, out}} = \int{\rho \omega^2 (u_x^2 + u_y^2)} \mathrm{d}s_{\text{in, out}}$. Here, $\omega$ represents the angular frequency, $s_{\text{in}}$ and $s_{\text{out}}$ denote the regions of the reference medium adjacent to the constrained structure. The mode conversion efficiency~\cite{kweun2017transmodal} is calculated similarly, with a conversion factor $\alpha = V_s / V_p$, where $V_p$ and $V_s$ are the velocities of the P- and S-modes respectively. Thus, the conversion rate for the P-to-S medium is represented as $\alpha T_{tr}$, and for the S-to-P medium, as $\alpha^{-1} T_{tr}$. Our spectral analysis reveals high transmission efficiencies in the harmonimode media, along with notable conversion rates in the mode conversion media, particularly within the 3-10 kHz range. This finding underscores the impressive broadband capabilities of our designs. However, the transmission spectrum shows a slight reduction in transmission or conversion efficiency at the frequency band’s extremities, likely due to resonance at higher frequencies and cutoff at lower frequencies.

In summary, we have developed and explored the elastodynamics of a constrained metamaterial, merging a classical mass-spring lattice approach with an innovative constrained design. Our design, which restricts specific degrees of freedom, demonstrates remarkable capabilities in harmonimode behavior for impedance matching and wave filtering, as well as the facile formation of functional gradients for perfect mode conversion and non-reciprocal wave propagation. Utilizing both continuum and microstructure-based perspectives, we've shown through 2D numerical simulations the unique wave propagation characteristics of these media. Moreover, we demonstrate that this constraint-based design approach can be effectively integrated with DTE, allowing for the simultaneous modulation of elastic wave paths and modes in a highly controlled manner. A pivotal aspect of our research is the practical realization of these concepts through a strategically designed hinged structure, which is both theoretically robust and practically feasible.

This work represents a new direction in metamaterial research, emphasizing the role of constraints in structural elements and paving the way for applications that extend well beyond wave steering and filtering. Future work will focus on the experimental validation of the proposed media and related functions for exploring the novel elastodynamic behaviors. Additionally, the development of tunable or active metamaterials by integrating constrained media with electromechanical systems presents an intriguing research direction.

G.L.H. acknowledge the Air Force Office of Scientific Research under Grant No. AF9550-18-1-0342 and AF 9550-20-0279 with Program Manager Dr. Byung-Lip (Les) Lee. Z. C. thanks the fund support from the 2115 Talent Development Program of China Agricultural University and Fundamental Research Funds for the Central Universities.

\bibliography{Reference}

\end{document}